\documentstyle[12pt]{article}
\begin{document}
\def\lb{\label}
\def\P{\Pi_{p+1}}
\def\t{\theta}
\def\be{\begin{equation}}
\def\ee{\end{equation}}
\def\d{\partial}
\def\a{\alpha}
\def\b{\beta}
\def\g{\gamma}
\def\G{{\cal G}}
\def\gg{{\bf g}}
\def\theequation{\thesection.\arabic{equation}}
\def\F{{\cal F}}
\def\H{{\cal H}}
\def\E{{\cal E}}
\def\e{\epsilon}
\def\l{\lambda}
\def\O{\Omega}
\def\o{\omega}
\def\D{{\cal D}}
\def\T{\Theta}
\def\ep{\varepsilon}
\def\tdt{\tilde{\theta}}
\def\tdz{\tilde{z}}
\def\tdD{\tilde{{\cal D}}}
\def\tdd{\tilde{\partial}}
\def\H{{\cal H}}
\def\bd{\bar{\d}}
\begin{titlepage}
\title{On Paragrassmann Differential Calculus}
\author{\em  A.T.Filippov
\thanks{Address until Dec.22, 1992: Yukawa Inst. Theor. Phys., Kyoto
Univ., Kyoto 606, Japan
 ~~~e-mail: filippov@jpnyitp.bitnet~~~
filippov@jpnyitp.yukawa.kyoto-u.ac.jp},
A.P. Isaev
\thanks{e-mail: isaev@theor.jinrc.dubna.su} and
A.B.Kurdikov
\thanks{e-mail: kurdikov@theor.jinrc.dubna.su}. \\
Laboratory of Theoretical Physics, JINR, Dubna, \\
SU-101 000 Moscow, Russia
}
\date{}
\maketitle

\begin{abstract}
This paper significantly extends and generalizes our previous paper
\cite{fik}. Here we discuss explicit general constructions
for paragrassmann calculus with one and many variables.
For one variable, nondegenerate differentiation algebras
are identified and shown to be equivalent to the algebra of
$(p+1) \times (p+1)$ complex matrices.
If $(p+1)$ is  prime integer, the algebra is nondegenerate  and so
unique.  We then give a general construction of the many-variable
differentiation algebras. Some particular examples are related to the
multi-parametric  quantum deformations of the harmonic oscillators.
\end{abstract}
\end{titlepage}
\newpage

\begin{center}
{\bf Dedication}
\end{center}
This paper is in memory of Mikhail Konstantinovich Polivanov.
One of the authors (A.T.F.) had a privilege to be a friend of
him for many and many years. He was not only a distinguished scientist
but a true Russian intellectual having deep roots in Russian culture.
It is a deep sorrow that we can no more have a talk with him on science,
poetry, religion\dots

\section{Introduction}
\setcounter{equation}0
Paragrassmann algebras (PGA) are interesting for several reasons. First,
they are relevant to conformal field theories \cite{zam}. Second, studies
of anyons and of topological field theories show the necessity of unusual
statistics. These include not only the well-known parastatistics but
fractional statistics as well (see, e.g. Refs.~\cite{for}).
One can also find in recent literature some hints \cite{pol} at a
connection between PGA and
quantum groups \cite{man}, \cite{jim}. Ref.~\cite{fik} has demonstrated
this connection in some detail, showing that some special PGA are closely
related to $q$-oscillators \cite{bied}, \cite{kul} and to a differential
calculus on quantum hyperplanes \cite{wz}, \cite{fz}.
Finally, it looks aesthetically appealing
to find a generalization of the Grassmann analysis \cite{ber} that proved
to be so successful in describing supersymmetry.

Recently, some applications of PGA have been discussed in literature.
In Ref.~\cite{rub} that inspired many other investigations, a
parasupersymmetric generalization of quantum mechanics had been proposed.
Ref.~\cite{dur} has attempted at a more systematic consideration
of the algebraic aspects of PGA based on the Green {\em ansatz} (see,
e.g. \cite{jpn}) and introduced, in that frame, a sort of paragrassmann
generalization of the conformal algebra.
Applications to the relativistic theory of the first-quantized spinning
particles have been discussed in \cite{kor}. Further references can be
found in \cite{fik}, \cite{spir}.

The aim of this paper is to construct a consistent generalization of the
Grassmann algebra (GA) to a paragrassmann one preserving, as much as
possible, those features of GA that were useful in physics applications.
A crucial point of our approach is defining generalized derivatives in
the paragrassmann variables satisfying natural restriction allowing to
construct a differential calculus. As in the previous paper \cite{fik},
here we mainly concentrate on the algebraic aspects leaving the
applications to future publications.

Section 2 treats algebras generated by one paragrassmann variable
$\theta$, $\theta^{p+1} =0$, and a differentiation operator $\d$.
This generalized differentiation coincides with the Grassmann one for
$p=1$ and with the standard differentiation
when $p\rightarrow \infty $. We construct a most general
realization of these algebras and identify a set of nondegenerate
ones which are proved to be equivalent. Simplest useful
realizations are presented in Section 3.

In Section 4, simplest PGA generated by many variables
$\theta_{i}$ and corresponding differentiations $\d_i$ are defined.
They obey the nilpotency condition $\theta^{p+1}=0$ ($\d^{p+1}=0$)
  where $\theta$ ($\d$) is any linear combination of $\theta_{i}$
 ($\d_i$), and appear to be naturally related to the
 non-commutative spaces satisfying the commutation relations
 $\theta_{i}  \theta_{j}=q_{ij} \theta_{j} \theta_{i}\;,\; i<j$
 (and similar relations for $\d_i \d_j$), where $q_{ij}^{p+1} =1$.
These relations once more demonstrate a deep connection between PGA and
quantum groups with deformation parameters $q$ being roots of unity.

Section 5 briefly summarizes the results and presents one more relation
of our algebras to quantum groups as well as a speculation on possible
applications.

\section{Differential Calculus with One Variable}
\setcounter{equation}0
In Ref.~\cite{fik} we have considered paragrassmann algebras
$\Gamma_{p+1}(N)$ with $N$ nilpotent variables $\theta_{n}$,
$\theta_{n}^{p+1}=0$, $n=1,\ldots,N$. Some wider algebras
$\Pi_{p+1}(N)$ generated by $\theta_{n}$ and additional
nilpotent generators $\d_{n}$ have also been constructed.
These additional generators served for defining a paragrassmann
differentiation and paragrassmann calculus. The building block
for this construction was the simplest algebra $\Pi_{p+1}(1)$.
By applying a generalized Leibniz rule for differentiations in
the paragrassmann algebra $\Gamma_{p+1}(N)$
we have found two distinct realizations for $\Pi_{p+1}(1)$ closely
related to the $q$-deformed oscillators.
We have mentioned in \cite{fik} that other realizations of the
$\Pi_{p+1}(1)$  may be constructed. The aim of this section is
to demonstrate this in detail. We shall also show that, under
certain conditions, all these realizations are equivalent and one
may choose those which are most convenient for particular problems.

 Intuitively, paragrassmann algebra $\P$ should be understood as
 some good $p$-generalization of the classical fermionic algebra
 $\Pi_{2}$
 \begin{eqnarray}
\t^{2} = 0 & = & \d^{2} \;\;, \\        \lb{u1}
 \d \t + \t \d & = & 1 \;\;.            \lb{u2}
 \end{eqnarray}
By `$p$-generalization' we mean that (\ref{u1}) is to be replaced by
\be
                                        \lb{u3}
\t^{p+1} = 0  =  \d^{p+1} \;\;,
\ee
(it is implied, of course, that $\t^{p}\not=0$ and the same for $\d$ ).
So the question is: which generalization of (\ref{u2})
might be called `good'.
Many variants have been tried already (see for example \cite{jpn}).
As a rule, they deal with certain symmetric multilinear
combinations, like
$\t^{2}\d+\t\d\t+\d\t^{2}$ (for $p=2$), and meet with
difficulties when commuting $\t$ and $\d$.

To find a correct generalization recall that Eq.~(\ref{u2}) allows to
define the Grassmann differential calculus. It shows how to push
the differentiation operator $\d$ to the right of the variable
$\theta$. On the other hand, representing $\d$ and $\theta$ by
$2 \times 2$ real matrices, we can make them Hermitian conjugate
and thus interpret as annihilation and creation operators.
Then Eq.~(\ref{u2}) is the normal-ordering rule. The second important
feature of this relation is that it preserves the Grassmann
grading, $-1$ for $\d$ and $+1$ for $\theta$. In physics terminology,
this means that the normal-ordering is not changing the number of
`particles'.

Thus, to construct a generalization of the relation (\ref{u2}),
we first define a natural grading in the associative algebra
generated by $\theta$ and $\d$ obeying Eq.~(\ref{u3}),
\be
\lb{u4}
deg\;(\t^{r_{1}}\d^{s_{1}}\t^{r_{2}}\d^{s_{2}}\dots
 \t^{r_{k}}\d^{s_{k}}\;) = \Sigma r_{i} - \Sigma s_{i}\;,
 \ee
and denote by $\P(l)$ the linear shell of monomials of the degree $l$.
Then our basic requirement is:
\be
\lb{u5}
a\;set\;L^{(l)}=\{\t^{r}\d^{s}\;,\;r-s=l\;\}\;\;forms\;a\;
basis\;of\;\P(l)\;.
\ee
This immediately reduces the range of possible degrees to
$0\leq l \leq p$ and makes all the subspaces $\P(l)$ and the
entire algebra
\be
\lb{u6}
\P = \oplus^{p}_{l=-p} \P(l)
\ee
finite-dimensional:
$$
\pi^{l} \equiv dim(\P(l)) = p+1-| l | \;\;,\;\;dim(\P) = (p+1)^{2}\;.
$$
Then, by applying the assumptions (\ref{u4}) and (\ref{u5})
to $\d \theta$, we find that
\be
\lb{u7}
\d \t = b_{0} + b_{1}\t \d +b_{2}\t^{2} \d^{2} + \dots +
b_{p}\t^{p} \d^{p},
\ee
where $b_i$ are complex numbers restricted by consistency of
 the conditions (\ref{u3}) and (\ref{u7}) and by further assumptions
to be formulated below. With the aid of Eq.~(\ref{u7}), any element
of the algebra can be expressed in terms of the basis
$\theta^r \d^s$, i.e. in the normal-ordered form.

A useful alternative set of parameters, $\a_{k}$, also fixing
the algebra may be defined by
\be
\lb{u8}
\d \t^{k} = \a_{k}\t^{k-1} + (\dots )\d\; ,
\ee
where the dots denote a polynomial in $\theta$ and $\d$. This relation is
a generalization of the commutation relation for the standard
derivative operator, $\d_z z^k = k z^{k-1} + z^k \d_z$, and we may
define the differentiation of powers of $\theta$ by analogy,
\be
\lb{pa14i}
\d (\t^{k}) = \a_{k}\t^{k-1} , \; \alpha_0 \equiv 0 \; ,
\ee
to be justified later.

By applying Eq.~(\ref{u7}) to Eq.~(\ref{u8}) one may derive the
recurrent relations connecting these two sets of the parameters:
\begin{eqnarray}
\a_{1} & = & b_{0} \; , \nonumber \\
\a_{2} & = & b_{0} + b_{1}\a_{1} \; , \nonumber \\
\a_{3} & = & b_{0} + b_{1}\a_{2} + b_{2}\a_{1}\a_{2} \; ,
 \lb{pa14} \\
 \;\; &...&  \nonumber \\
\a_{k+1} & = & \sum^{k}_{i=0}
b_{i} \frac{(\a_{k})!}{(\a_{k-i})!} \; ,\nonumber \\
 \;\; &...&  \nonumber
\end{eqnarray}
where $(\a_{k})! \equiv \a_{1}\a_{2} \cdots \a_{k}$.
These relations enable us to express $\a_{k}$ as a function of
the numbers $b_{i}\;,\; 0 \leq i \leq k-1$ .
The first few expressions are
\begin{eqnarray}
\a_{1} & = & b_{0}, \;\; \a_{2} = b_{0}\frac{1-b_{1}^{2}}{1-b_{1}} , \;\;
\a_{3} = b_{0}\frac{1-b_{1}^{3}}{1-b_{1}} +b_{2}b_{0}^{2}(1+b_{1}),
\nonumber  \\
\a_{4} & = & b_{0}\frac{1-b_{1}^{4}}{1-b_{1}} +b_{2}b_{1}b_{0}^{2}
(1+b_{1})+ b_{0}(b_{3}+b_{2}b_{0})(1+b_{1})\a_{3}, \; \dots \nonumber \\
\end{eqnarray}
The inverse  operation, deriving $b_{i}$ in terms of $\a_{k}$,
is well-defined only if all $\a_{k}\not=0$.

The consistency condition mentioned above requires that the parameters
must be chosen so as to satisfy the identity
 $$0 \equiv \d \theta^{p+1}.$$
Taking into account that the second term in Eq.~(\ref{u8})
vanishes for $k= p+1$ we have $\a_{p+1}= 0$, with no other restrictions
on the parameters $\alpha_k$ with $k \leq p$. The corresponding
restriction on $p+1$ parameters $b_i$ follow  from Eq.~(\ref{pa14}),
\be
\lb{u9}
\a_{p+1}(b_{0},\dots ,b_{p})\equiv
 b_{0} + b_{1}\a_{p} + b_{2}\a_{p}\a_{p-1} + \dots +
b_{p}\a_{p}\a_{p-1} \cdots \a_{2}\a_{1} =0 \; ,
\ee
where the parameters $\alpha_i$ are expressed in terms of $b_i$.
Any admissible set $\{b\}$ determines an algebra $\P^{\{b\} }$
with the defining relations (\ref{u3}), (\ref{u7}).
To each algebra $\P^{\{b\} }$ there corresponds a set $\{\a\}$.
{\it A priori}, there are no restrictions on $\{\a\}$
but, if we wish to treat $\d$ as a non-degenerate derivative
with respect to $\t$, it is reasonable to require,
in addition to (\ref{u5}), that
\be
\lb{u10}
all \;\;\a_{k} \not= 0\;.
\ee
So let us call a set $\{b\}$ (and corresponding algebra$\P^{\{b\} }$)
{\it non-degenerate}, if the condition (\ref{u10}) is fulfilled,
and degenerate otherwise.
As it was already mentioned, in the non-degenerate case the numbers
$b_{i}$ are completely determined by the numbers $\a_{k}$,
so we can use the symbol $\{\a\}$ as well as $\{b\}$.

In general, different sets $\{b\}$ determine non-equivalent algebras
$\P^{\{b\} }$. At first sight,
the algebras corresponding to different sets $\{b\}$  look very
dissimilar. However, this is not true for the non-degenerate ones.
In fact, all non-degenerate algebras $\P^{\{b\} }$ are isomorphic to
the associative algebra $Mat(p+1)$ of the complex
$(p+1) \times (p+1)$ matrices.

This isomorphism can be manifested by constructing an explicit
exact (`fundamental') representation for $\P^{\{b\}}$ . With this
aim, we treat $\t$ and $\d$ as creation and annihilation operators
(in general, not Hermitian conjugate) and introduce the ladder of
$p+1$ states $|k\rangle $, $k=0,1, \ldots ,p$ defined by
\be
\lb{pa8}
\d |0\rangle = 0 \;, \; |k\rangle \sim \t^{k} |0\rangle \; ,
\;\;  \t |k\rangle = \beta_{k+1} |k+1\rangle \; .
\ee
Here $\beta_k$ are some non-zero numbers, reflecting the freedom of
the basis choice. As $|p+1\rangle =0$, the linear shell of
the vectors $|k\rangle$ is finite-dimensional
 and in the nondegenerate case, when all $\beta_{k} \neq 0 \;\;\;
 (k=1, \dots ,p)$, its dimension is $p+1$.

Using (\ref{pa8}) and (\ref{u8}) we find
\be
\lb{pa11}
\d |k\rangle = (\a_{k} /\beta_{k}) |k-1\rangle \;\; .
\ee
Thus the fundamental (Fock-space) representations of the operators
$\t$ and $\d$ is
\begin{eqnarray}
\lb{pa13a}
\t_{mn} & = & \langle m| \t |n\rangle =\beta_{n+1} \delta_{m,n+1}\;\; , \\
\d_{mn} & = & \langle m| \d |n\rangle =(\a_{n} / \beta_{n} )
\delta_{m,n-1} \;\; .
\lb{pa13b}
\end{eqnarray}
It is not hard to see that, for non-zero parameters $\a$,
the matrices corresponding to $\t^{m}\d^{n}\;\;(m,n=0\dots p)$
form a complete basis of the algebra $Mat(p+1)$. The isomorphism
is established.

Nothing similar occurs for degenerate algebras.
To show an evidence against using them in the paragrassmann
calculus, consider an extremely degenerate algebra with
$\a_{k}=0$, so that $b_{0}=b_{2}=\dots =b_{p}=0\;,\;b_{1}\not=0$.
This algebra has nothing to do with $Mat(p+1)$, and its properties
essentially depend on the value of $b_{1}$.
It is abelian if $b_{1}=1$; it is a paragrassmann algebra
of the type $\Gamma_{p+1}(2)$ if $b_{1}$ is a primitive root
of unity (see \cite{fik}), and so on. We hope this remark is not
sounding like a death sentence on the degenerate algebras.
At least, it has to be suspended until further investigation
which will probably prove their usefulness in other contexts.
However, if we wish to have paragrassmann calculus similar to
the Grassmann one, we have to use the nondegenerate algebras.

In  Ref.~\cite{fik}, we mentioned that the structure of the algebra
$\Gamma_{p+1}$ depend on  arithmetic  properties  of $p+1$ and
that this  may give certain restrictions on algebras $\Pi_{p+1}$.
Here we can make a much stronger statement. Using Eqs.
(10) it is easy to prove that if $\alpha_{n+1}=0$ for some
positive $n<p$, then $n+1$ is  a divisor of  $p+1$.
It follows that for prime integer values of $p+1$ the condition
(13) is satisfied if $\alpha_1 \ne 0$ (for $\alpha_1 =0$ all
$\alpha_n$ must vanish). Thus all nontrivial algebras $\Pi_{p+1}$
are nondegenerate for prime integer $p+1$.
It  also follows that nonequivalent degenerate algebras $\Pi_{p+1}$
can be classified by  divisors  of  $p+1$ in the following sense.
Let $n+1$ is a divisor of $p+1$, $\alpha_{n+1} = 0$, $\alpha_m \ne 0$
for $1 \le m \le n$. Then all $\alpha_m$ with $m > n+1$ are defined
by the obvious periodicity condition, $\a_{m+ n+1} = \a_m$, and $b_m$
with $0 \le m \le n$ are expressed in terms of $\a_m$. It is easy to
show that $\d^{n+1} = 0$ and so the commutation relation (7) is also
completely fixed. This analysis can be pushed further but in what
follows we restrict our consideration to the nondegenerate algebras.

Thus, two natural requirements (\ref{u5}) and (\ref{u10})
reduce the range of possible generalizations of the
fermionic algebra $\Pi_{2}$ to the unique algebra $\P$ that is
isomorphic to $Mat(p+1)$~\footnote{H.~Weyl in his famous book
on quantum mechanics had foreseen relevance of these algebras
to physics problems. After detailed description of the spin
algebras he discussed more general finite algebras and remarked that
the finite algebras like those discussed here will possibly appear in
future physics. We think it natural to call $\P$ the `finite Weyl
algebra' or `para-Weyl algebra'.}.
The grading (\ref{u4}) in $\P$ corresponds to
`along-diagonal' grading in $Mat(p+1)$.
Different non-degenerate algebras  $\P^{\{b\} }$
are nothing more than alternative ways of writing
one and the same algebra $\P$.  We will call them {\it versions}
having in mind that fixing the $b$-parameters is analogous to
a gauge-fixing (in H.~Weyl's usage).

This implies that we will mainly be interested in `version-covariant'
results, i.e. independent on a version choice.  Nevertheless, special
versions may have certain nice individual features making them
more convenient for concrete calculations (thus allowing for simpler
derivations of covariant results by non-covariant methods).
Several useful versions will be described below.
Before turning to this task we end our general discussion with
several remarks.

{\em First}. The existence of the exact matrix representation
(\ref{pa13a}), (\ref{pa13b}) is very useful for deriving
version-covariant identities in the algebra $\P$. For instance,
it is easy to check that
\begin{eqnarray}
\{ \d\;,\;\t^{(p)}\} & = &
\left(\Sigma \a_{k}\right)\;\t^{p-1}\;,       \lb{u12}
\cr
\{ \d^{p}\;,\;\t^{(p)}\} & = & \Pi \a_{k}\;,      \lb{u13}
\end{eqnarray}
and to find many other relations. Here we have introduced a useful
notation
\be
\lb{u14}
\{ \Xi \; ,\;\; \Psi^{(l)} \} = \Xi \Psi^{l} + \Psi \Xi \Psi^{l-1}
+ \dots + \Psi^{l} \Xi \; .
\ee
The identities (\ref{u12}) generalize those known in the
parasupersymmetric quantum mechanics \cite{rub}.

{\em Second}. One may adjust the parameters  $\beta_{k}$
to get a convenient matrix representation for $\t$ and $\d$.
As a rule, we take $\beta_{k} =1$. Note that
for the versions with real parameters $\a_{k}$, it is possible to choose
$\beta_{k}$ so as to have $\t^{\dag}=\d$ .
We also normalize $\t$ and $\d$ so that $\a_{1}\equiv b_{0}=1$. This
gives a more close correspondence with the Grassmann relation (\ref{u2}).

{\em Third}. In a given (non-degenerate) version $\P^{\{b\}}$
the components of the vector
$R_{\{b\}}^{(l)}=col\{\d^{j}\t^{i} \}_{i-j=l}$
form a basis of the subspace $\P(l)$ that is completely equivalent
to the original one having the components
$L_{\{b\}}^{(l)}=col\{\t^{i}\d^{j} \}_{i-j=l} \;$, see Eq.~(\ref{u5}).
Hence, there must exist a non-degenerate matrix
$C^{(l)}_{\{b\}} \in mat(\pi^{(l)} \; , \; \bf C )$
connecting these two bases,
\be
\lb{pa7}
R_{\{b\}}^{(l)} = C^{(l)}_{\{b\}} \cdot L_{\{b\}}^{(l)} \; ,\;\;
l=-p,\dots , p\;.
\ee
The elements of the $C$-matrix are certain functions of $b_{i}$
which are usually not easy to calculate except simple versions.
The original commutation relation (\ref{u7})
is also included in the system (\ref{pa7}), for $l=0$.

Quite similarly, two $L$-bases ($R$-bases) taken in different versions
$\{b\}$ and $\{b'\}$ are connected by a non-degenerate matrix
$M_{\{bb'\}}$ ($N_{\{bb'\}}$), i.e.
\be
L_{\{b\}} =M_{\{bb'\}}L_{\{b'\}} \; ,
\lb{pa18}
\ee
\be
R_{\{b\}} =N_{\{bb'\}}R_{\{b'\}} \; ,
\lb{pa19}
\ee
where the indices $(l)$ are suppressed.
The matrices $M_{\{bb'\}}^{(l)}$ (and $N_{\{bb'\}}^{(l)}$)
belong to $Mat(\pi^{(l)})$ and obey the cocycle relations:
$$
M_{\{bb'\}}M_{\{b'b\}}=1 \; , \;\;
M_{\{bb'\}}M_{\{b'b''\}}M_{\{b''b\}}=1 \; .
$$
By applying Eq.~(\ref{pa7}), we immediately get the relation
\be
N_{\{bb'\}}=C_{\{b\}}M_{\{bb'\}}C^{-1}_{\{b'\}} \;
\lb{pa20}
\ee
that permits evaluating $C$-matrices for complicated versions once
we know them in one version.
In particular, Eq.~(\ref{pa18}) tells that the operator
$\d$ in any version can be represented as a linear combination
of the operators
$\d \; , \;\; \t\d^{2} \; , \;\; \dots , \;\;\t^{p-1}\d^{p} $
of any other version.  We shall  soon see
that this, for instance, allows to realize $q$-oscillators
in terms of generators $\t$ and $\d$ of other versions, and
{\it vice versa}.

\section{Versions of the Paragrassmann Calculus}
\setcounter{equation}0
Now consider some special versions related to the simplest
forms of Eq.~(\ref{u7}).

(0): {\bf Primitive Version} \\
Here $  b_{1}= \dots =b_{p-1}=0 \; ,\;\;
b_{p}=-1 \; , \;\; {\rm so \;\; that} \;\; \a_{i} =1 \; ,
$
\be
(\d_{(0)})_{mn}= \delta_{m,n-1} \; , \;\; \d_{(0)}
\t =1-\t^{p}\d^{p}_{(0)} \; .
\lb{pa15}
\ee
This realization of $\d$ may be called `almost-inverse' to $\t$.
In the matrix representation (\ref{pa13a}), (\ref{pa13b})
with $\beta_{k} =1$ we have $\t^{T}=\d_{(0)}$.
This version is the simplest possible but the differential calculus
is a fancy-looking thing in this disguise and it is unsuitable
for many applications. Still, it has been used in some applications.
For example, the operators $\t$ and $\d_{(0)}$ for $p=2$ coincide with
parafermions in the formulation of the parasupersymmetric quantum
mechanics \cite{rub}.

(1): {\bf q-Version, or Fractional Version} \\
Here $ b_{1}=q \neq 0 \; , \;\; b_{2}=b_{3}= \dots =b_{p}=0 \; ,
\; {\rm so \;\; that}$
$$
\a_{i}=1+q+ \dots +q^{i-1} = \frac{1-q^{i}}{1-q} \; .
$$
The condition $\alpha_{p+1}=0$ tells that $q^{p+1}=1 \;\; (q \neq 1)$,
while the assumption that all $\a_{i} \neq 0$ forces $q=b_1$ to be
a {\em primitive} root, i.e. $q^{n+1} \neq 1 \; , \;\; n<p$ . Thus in
this version ($\d =\d_{(1)}$)
\be
\d_{(1)} \t = 1+ q\t \d_{(1)} \; ,
\lb{pa16}
\ee
$$
\d_{(1)}(\t^{n})=(n)_{q} \t^{n-1}\; , \;\;(n)_{q}=\frac{1-q^{n}}{1-q} \; .
$$
These relations were derived in Ref.~\cite{fik} by assuming that
$\d$ is a generalized differentiation operator, i.e.
satisfying a generalized Leibniz rule (a further generalization is
introduced below). The derivative $\d_{(1)}$ is naturally related to
the $q$-oscillators ($q$-derivative) and to quantum algebras
(see \cite{fik} and references therein).
Eq.~(\ref{pa16}) is also extremely convenient for
 generalizing to Paragrassmann algebras with many $\t$ and $\d$.

(2): {\bf Almost Bosonic Version} \\
For this Version
$$
 b_{1}=1 \; , \;\; b_{2}= \dots =b_{p-1}=0 \; , \;\;
b_{p} \neq 0 \; , \;\; {\rm so \;\; that} \;\; \a_{k}=k \; .
$$
The condition $\alpha_{p+1}=0$ gives $b_{p}=-\frac{p+1}{p!} \;$ and thus
\be
(\d_{(2)})_{mn}=n\;\delta_{m,n-1} \; , \;\;
\d_{(2)} \t = 1+ \t \d_{(2)} -\frac{p+1}{p!}\; \t^{p} \d_{(2)}^{p} \; .
\lb{pa17}
\ee
As $\d_{(2)}(\t^{n}) = n \t^{n-1} \;\; (n \neq p+1 ) \; $,
this derivative is `almost bosonic'.

Let us now discuss the interrelations between $\t$ and $\d$.
As we have already mentioned, the notation itself hints at treating
$\d$ as a derivative with respect to $\t$ (see (\ref{u8})).
To be more precise,
let us represent vectors as functions of $\theta$
$$
|F \rangle = \sum_{k=0}^{p} f_{k} | k \rangle \;\; \Leftrightarrow
F (\t) = \sum_{k=0}^{p} f_{k} \t^{k}.
$$
The action of the derivative $\d$ on this function
is defined by (\ref{pa8}) and (\ref{pa11}) ($\beta_{k} =1$),
\be
\lb{pa21a}
\d (1)=0 \; ,\;\; \d (\t^{n})= \a_{n}\t^{n-1} \;\; (1 \leq n \leq p) \; .
\ee
It is however clear that this derivative does not
obey the standard Leibniz rule $\d(ab)=\d(a)b + a\d(b)$.

So consider the following  modification of the Leibniz rule \cite{fik},
\cite{liealg}
\be
\lb{pa22}
\d(FG) = \d(F) \bar{\gg} ( G) + \gg (F) \d(G) \; .
\ee
The associativity condition (for differentiating $FGH$) tells that
$\gg$ and $\bar{\gg}$ are homomorphisms, i.e.
\be
\lb{pa23}
\gg(FG)=\gg(F)\gg(G) \; , \;\;
\bar{\gg}(FG)=\bar{\gg}(F)\bar{\gg}(G) \; .
\ee
The simplest natural homomorphisms compatible with the relations
(\ref{pa21a}), (\ref{pa22}) and (\ref{pa23}) are linear automorphisms
of the algebra  $\Gamma_{p+1}$,
\be
\lb{pa24}
\gg(\t)=\gamma \t \; , \;\;
\bar{\gg}(\t)=\bar{\gamma} \t \; ,
\ee
where $\gamma \; , \; \bar{\gamma}$  are arbitrary
complex parameters and
\be
\lb{pa25}
\a_{k} = \frac{{\bar{\gamma}}^{k} - \gamma^{k}}{\bar{\gamma} - \gamma} .
\ee
Now the condition (\ref{u9}) yields the equation
\be
\lb{pa26}
\a_{p+1} =
\frac{{\bar{\gamma}}^{p+1} - \gamma^{p+1}}
{\bar{\gamma} - \gamma} =0 ,
\ee
and assuming nondegeneracy requirements $\a_{n} \neq 0 \; (n < p+1)$
we conclude that $\bar{\gamma}/\gamma $ must be a primitive $(p+1)$-root
of unity. Thus we may formulate another interesting version
of the paragrassmann algebra $\P$.

(3):{\bf $\gg-\bar{\gg}$--Version} \\
In this version, the parameters $\alpha_k$ are supposed to be given
by Eq.~(\ref{pa25}) and we
can calculate $b_i$ by solving Eq.~(\ref{pa14}):
$$b_{0}=1, \; b_{1} = \bar{\gamma} + \gamma -1, \;
b_{2} = (\bar{\gamma} -\bar{\gamma}\gamma + \gamma -1)/
(\bar{\gamma} + \gamma), \; \dots $$
Here $\gamma$ and $\bar{\gamma}$ are complex numbers constrained only by
the condition that $q=\bar{\gamma}/\gamma$ is a primitive root of unity
$$
\left( \frac{\bar{\gamma}}{\gamma} \right)^{p+1} =1 .
$$
 From Eqs.~(\ref{pa22}) and (\ref{pa24}) one can derive the following
operator relations for the automorphisms $\gg, \; \bar{\gg}$
\be
\lb{pa27}
\d \t-\gamma \t \d = \bar{\gg} \; , \;\; \d \t-\bar{\gamma}\t \d = \gg .
\ee

For the special case
$\gamma = (\bar{\gamma})^{-1} = q^{1/2}$, redefining
$\d = a, \; \t = a^{\dag}$
allows to recognize in (\ref{pa27}) the definitions of the
$q$-deformed oscillators in the Biedenharn-MacFarlane form \cite{bied}.
Note that Version-(1) can be derived  from Version-(3) by putting
$\bar{\gamma}=q \; \gamma=1$, (or, $\bar{\gamma}=1 \; \gamma=q$).
So we may regard Version-(3) as a generalization of
Version-(1). Moreover, it can be shown that, for $p=2$, both Version-(0)
and Version-(2) are specializations of Version-(3).
However, it is not true for $p > 2$ and, in general, the Leibniz rule
(\ref{pa22}) has to be further modified.
To find a most general deformed Leibniz rule we slightly
change the definition of the $\gg-\bar{\gg}$--Version.

(4): {\bf Generalized Version} \\
Namely, leaving untouched the equations
(\ref{pa21a}) and (\ref{pa25}), we assume that $\gamma$ and
$\bar{\gamma}$ are arbitrary parameters not constrained by
Eq.~(\ref{pa26}), i.e.
\be
\lb{pa28}
\tilde{\a}_{p+1} \equiv \frac{{\bar{\gamma}}^{p+1} - \gamma^{p+1}}
{\bar{\gamma}-\gamma} \neq 0 \;\;\; {\rm but} \;\; \a_{p+1}=0 .
\ee
Then the conditions (\ref{u9}), (\ref{pa21a}), and (\ref{pa25}) are
only fulfilled if the equations (\ref{pa27}) are modified as follows:
\be
\lb{pa29}
\begin{array}{rcl}
\d \t - \gamma \t \d & = & \bar{\gg}-\frac{\tilde{\a}_{p+1}}{(\a_{p})!}
\t^{p}\d^{p} \; , \\ \\
\d \t - \bar{\gamma} \t \d & = & \gg-\frac{\tilde{\a}_{p+1}}{(\a_{p})!}
\t^{p}\d^{p} \; .
\end{array}
\ee
Version-(0) may be derived  from this version by substituting
$\gg(\t^{k})=\delta_{k,0}$ and $\bar{\gg}=1$; this means that
$\gamma =0, \; \bar{\gamma}=1$. Equivalently, we may choose $\gg=1$ and
$\bar{\gg}(\t^{k})= \delta_{k,0}$; then $\gamma =1, \; \bar{\gamma}=0$.
Versions (1) and (3) are reproduced if we put $\tilde{\a}_{p+1}=0$,
while Version-(2) may be obtained in the limit
$\gamma= \bar{\gamma} \rightarrow 1$. Thus, Version-(4)
generalizes all versions defined above.

The relations (\ref{pa29}) dictate a more general modification
of the Leibniz rule
\be
\lb{pa30}
\d(FG) = \d(F) \bar{\gg} ( G) + \gg (F) \d(G) + Lz(F,G) \; .
\ee
As follows  from Eqs. (\ref{pa29}), the additional term
$Lz(.\,,.)$ belongs to the one dimensional space $\{ | p \rangle \}$.
We suggest to call this term  the `Leibnizean'. Note that
the associativity condition for the rule (\ref{pa30}) requires
Eqs.~(\ref{pa23}) and the additional relation
$$
Lz(FG,H) + Lz(F,G)\bar{\gg}(H) = Lz(F,GH) + \gg(F)Lz(G,H) \; .
$$

Versions (1) and (2) evidently reproduce the Grassmann calculus
for $p=1$ while the limit $p \rightarrow \infty $
gives $\gg=\bar{\gg}=1$ and $Lz=0$, thus reproducing the standard
calculus in the dimension one. Other versions obeying the conditions
$\lim_{p \rightarrow \infty} (Lz) = 0$ and
$\lim_{p \rightarrow \infty} (\gg, \; \bar{\gg}) = 1$
are much more complicated
(e.g. $b_{1}= 1, \; b_{2}= \dots =b_{p-k} =0 \; ,
b_{p-k+1} \neq 0 \; , \dots \; , \; b_{p} \neq 0 \;$,
for some fixed $k \geq 2$).

Summarizing this discussion, we note that in constructing a paragrassmann
calculus for many variables we wish to have a generalized Leibniz rule.
A most natural generalization must look like
\be
\lb{leibniz}
\d_i (FG)= \d_j (F) \bar{\gg}_{i}^{j}(G)+\gg_{i}^{j}(F) \d_{j} (G) \; ,
\ee
where $\bar{\gg}$ and $\gg$ are some automorphisms and the summation over
$j$ is understood. Only Versions (1) and (2) seem to be suitable in this
context.

\section{Paragrassmann Algebras with Many Variables}
\setcounter{equation}0
Here we present explicit realizations of some paragrassmann
algebras $\P(N)$ generated by $N$ coordinates
$\t_{i}$, $\t_{i}^{p+1}=0$ ($i=1, \dots , N$),
and $N$ corresponding derivatives $\d_{i}\; , \d_{i}^{p+1}=0$.
The simplest (bilinear) algebras can be constructed in Version-(1).
Thus consider the algebra $\P(1)$ defined by
\be
\lb{m4}
\d\t-q\t\d=1 \; ,\;\; \d^{p+1}=\t^{p+1}=0 \; ,
\ee
where $q$ is any primitive $(p+1)$-root of unity
The algebra (\ref{m4}) was the starting point for
considering the fractional para-supersymmetry \cite{lec}.
Our motivation for using this version is its extreme simplicity.
Furthermore, it gives bilinear commutation relations for generators
of $\P(N)$ that are closely related to the definitions of the quantum
hyper-plane \cite{man},
covariant q-deformed oscillators \cite{kul}, and
differential calculus on the quantum hyperplane \cite{wz}.
Other versions given by Eq.~(\ref{u7}) can be considered similarly
but they yield non-bilinear multi-paragrassmann algebras (a generic
example will be given below).

The automprphsim generator emerging in the generalized Leibniz rule
can be written as
\be
\lb{m5}
\gg = \d\t-\t\d .
\ee
It is easy to check that
$$
\t\d =(\gg-1)/(q-1) \; ,\;\; \d\t = (q\gg-1)/(q-1) \; ,
$$
and
\be
\lb{m6}
\gg\t =q\t \gg \; ,\;\; \gg\d = q^{-1} \d \gg \; .
\ee
Using this operator we define $N$ paragrassmann variables
\be
\lb{m7}
\t_{i}=\gg^{\rho^{i}_{N}}\otimes \gg^{\rho^{i}_{N-1}} \otimes \cdots
\otimes \gg^{\rho^{i}_{i+1}} \otimes
\t \gg^{\rho^{i}_{i}}   \otimes \gg^{\rho^{i}_{i-1}}\otimes
\cdots \otimes \gg^{\rho^{i}_{1}}
\ee
with the obvious commutation relations
\be
\lb{m8}
\t_{i}\t_{j} = q^{\rho_{ij}} \t_{j}\t_{i} \;,\;\;i<j\;.
\ee

We wish to restrict $N(N-1)/2$ numbers
$\rho_{ij} =\rho^{i}_{j}-\rho^{j}_{i}$
so as any linear combination of $\t_{i}$ is nilpotent,
\be
\lb{m9}
(\sum_{i=0}^{p} c_{i}\t_{i})^{p+1}=0 \; ,
\ee
and hence $\t_{i}$ generate a paragrassmann algebra $\Gamma_{p+1}(N)$.
One simple choice is
\be
\lb{strip}
\rho_{ij}=a_{j} \; (i < j)\;,
\ee
with all $q^{a_{i}}$ being
primitive roots of unity. With this choice, all $\t_{i}$ for $i<j$
acquire the same multiplier $q^{a_j}$ in commuting through $\t_j$.
So, if Eq.~(\ref{m9}) is valid for the linear combinations of the
first $(j-1)$ thetas, we may apply Eqs.~(35), (36) of Ref.~\cite{fik}
and thus prove it to be valid for any number of thetas (provided that
all $q^{a_{i}}$ are primitive roots of unity).

The  {\em ansatz} (\ref{m7}) generalizes
the expressions for many thetas obtained in \cite{fik} by
certain recurrent procedure.
It is natural to define the derivatives $\d_{i}$ by
\be
\lb{m10}
\d_{i}=\gg^{\sigma^{i}_{N}}\otimes \gg^{\sigma^{i}_{N-1}} \otimes \cdots
\otimes \gg^{\sigma^{i}_{i+1}} \otimes
\gg^{\sigma^{i}_{i}} \d \otimes \gg^{\sigma^{i}_{i-1}}\otimes
\cdots \gg^{\sigma^{i}_{1}} .
\ee
Then the commutation relations for $\d_{i}$ are ($i<j$):
\begin{eqnarray}
\d_{i}\d_{j} & = & q^{\sigma_{ij}}\d_{j}\d_{i} \; , \;\; \sigma_{ij} =
\sigma^{j}_{i}-\sigma^{i}_{j} \; , \lb{m11} \\
\t_{i}\d_{j} & = & q^{-\sigma^{j}_{i} -\rho^{i}_{j}}\d_{j}\t_{i} \;  ,
\lb{m12} \\
\t_{j}\d_{i} & = & q^{-\sigma^{i}_{j} -\rho^{j}_{i}}\d_{i}\t_{j} \; .
\lb{m13}
\end{eqnarray}
Here the parameters $\sigma_{i}^{j}$ are to be chosen
so that any linear combinations of the derivatives $\d$ is also nilpotent
\be
 \lb{m13a}
(\sum_{i=0}^{p} c_{i}\d_{i})^{p+1}=0.
\ee

Now, to obtain a closed algebra with quadratic commutation relations
we have to solve the following {\em problem $(\star)$}:

{\em to express $\d_{i}\t_{i}$ as a linear combination of
$1$ and $\t_{j}\d_{j}\;,\;i,j=1,\dots,N$ .}

It is more convenient to deal with the expressions
\be
\lb{m14}
\d_{i}\t_{i}-q^{\tau^{i}_{i}+1} \t_{i}\d_{i} =
\gg^{\tau^{i}_{N}} \otimes \cdots
\otimes  \gg^{\tau^{i}_{i}} \otimes \cdots
\otimes \gg^{\tau^{i}_{1}}\equiv
\gg^{\{\tau^{i}\}} \; ,
\ee

\be
\lb{m15a}
\d_{i}\t_{i}-q^{\tau^{i}_{i}} \t_{i}\d_{i} =
\gg^{\tau^{i}_{N}} \otimes \cdots
\otimes  \gg^{\tau^{i}_{i}+1} \otimes \cdots
\otimes  \gg^{\tau^{i}_{1}} \equiv
\gg^{\{\tau^{i}\}_{+}} \; ,
\ee
where $\tau^i_j = \rho^i_j + \sigma^i_j .$
 The terms $\t_{j}\d_{j}$ can be represented in the form
\be
\lb{m15}
E_{j}\equiv (q-1)q^{\tau^{j}_{j}} \t_{j}\d_{j} =
\gg^{\tau^{j}_{N}}
\otimes  \cdots \otimes  \gg^{\tau^{j}_{j}}(g-1) \otimes
\cdots \otimes \gg^{\tau^{j}_{1}} =
\gg^{\{\tau^{i}\}_{+}} - \gg^{\{\tau^{i}\}} \; .
\ee
 It is not hard to realize that the {\em problem} ($\star$) is solvable
 if and only if, for any $i$, there exists a sequence of operators
(\ref{m15}) producing (\ref{m14}) or (\ref{m15a})  from
 $1\equiv \gg^{\{0\}}$. To formulate this more rigorously, consider
 $2N$ points  $\{\;\{\tau^{i}\}\;,\; \{\tau^{i}\}_{+}\;,i=1\dots N\;\}$
and $N$ oriented segments $E_{i}=\{\tau^{i}\}\rightarrow \{\tau^{i}\}_{+}$
in an $N$-dimensional space.
This set of data composes an oriented graph $\G$ that obviously
does not contain cycles since all the segments are mutually orthogonal.
After these preliminaries, we can formulate the following

C r i t e r i o n : \em The problem $(\star)$ is solvable
if and only if the correspondent graph $\G$ is connected
(and therefore an oriented tree) and contains the point \rm \{0\}.

In other words, this means that one can define an equivalence
relation $\sim$ on the set ${\cal T}=\{\{0\}\;,\; \{\tau^{i}\}\;,\;
\{\tau^{i}\}_{+}\;,i=1\dots N\;\}$, so that
\be
\lb{cri1}
\begin{array}{lrl}
a) & \; \{\tau^{i}\}\sim \{\tau^{i}\}_{+}\;, & i=1\dots N\;,\\
b) & u=v \Rightarrow u \sim v \;, & \forall u,v \in {\cal T}\;.
\end{array}
\ee
Then the criterion tells that the entire ${\cal T}$ must be a single
equivalence class.

This criterion gives a simple procedure for getting the commutation
relations of $\d_{i}$ and $\t_{i}$:

1. Draw an oriented tree with a root \{0\} and $N$ edges;

2. Label the edges by the numbers  from 1 to $N$;

3. Find a path  from \{0\} to the beginning of the $i$-th edge;

4. Moving along this path, write
$$
\gg^{\{\tau^{i}\}}=1\pm E_{j_{1}}\pm E_{j_{2}}\dots  \;,
$$
taking `$+$', if the move agrees with the orientation of the edge $j_{a}$
and `$-$'otherwise;

5. Use the expressions (\ref{m14}) and (\ref{m15}).

This algorithm exhausts all admissible possibilities.
In particular, it proves that all the numbers $\tau^{i}_{j}$ can
only b $0$ or $\pm 1$. Thus, it brings some restrictions on the
exponentials $\rho^{i}_{j}$ and $\sigma^{i}_{j}$, though not too strong.
There are no direct restrictions on the values of $\rho_{ij}$ or
relations between them coming  from the criterion.
So the last string of the algebra, the commutation
relations of $\d_{i}$ and $\t_{i}$, is almost independent of the
first four ones,
(\ref{m8}), (\ref{m11} -- \ref{m13}).
Note that algebras corresponding to the different graphs are
non-equivalent, at least at the level of linear combinations.

Let us now present two simplest examples of the paragrassmann
algebras $\P(N)$.

\be
\lb{m16}
\begin{array}{ll}
1). \;\;\;\;\;\;\;  & \tau^{i}_{j} = \rho^{i}_{j} + \sigma{i}_{j} = 0
\;\; (i \neq j)\; ,\cr
 & \tau^{i}_{i} = 0 \; , \; - 1 \;,
\end{array}
\ee
or, shortly, $\{\tau^{i}\}=\{0\}$ or $\{\tau^{i}\}_{+}=\{0\}$.
With this choice, the algebra is ($i<j$):
\be
\lb{m17}
\begin{array}{rcl}
\t_{i}\t_{j} & = & q^{\rho_{ij}} \t_{j}\t_{i}    \; , \\
\d_{i}\d_{j} & = & q^{\rho_{ij}} \d_{j}\d_{i}    \; ,     \\
\t_{i}\d_{j} & = & q^{-\rho_{ij}}  \d_{j}\t_{i}  \; ,   \\
\t_{j}\d_{i} & = & q^{\rho_{ij}} \d_{i}\t_{j}    \; , \\
\d_{i}\t_{i} - q^{2\tau^{i}_{i} +1} \t_{i}\d_{i} & = & 1 \; .
\end{array}
\ee

This algebra has been discussed in \cite{fik}, \cite{fz}.
The correspondent graph $\G$ is a bunch of $N$ segments coming  from
(or to) zero point.

\be
\lb{m22}
\begin{array}{ll}
2). \;\;\;\;\;\;\; & \tau^{i}_{j} = 1 \; , \;\; j < i \; ;\cr
                   & \tau^{i}_{j} = 0 \; ,\;\; j \geq i \; ,
\end{array}
\ee
or, shortly, $\{\tau^{1}\}=\{0\}\;,\;\{\tau^{i+1}\}=\{\tau^{i}\}_{+}$.
Here we obtain ($i<j$):
\be
\lb{m23}
\begin{array}{rcl}
\t_{i}\t_{j} & = & q^{\rho_{ij}} \t_{j}\t_{i}    \; , \\
\d_{i}\d_{j} & = & q^{\rho_{ij} +1} \d_{j}\d_{i}    \; ,     \\
\t_{i}\d_{j} & = & q^{-1-\rho_{ij}}  \d_{j}\t_{i}  \; ,   \\
\t_{j}\d_{i} & = & q^{\rho_{ij}} \d_{i}\t_{j}    \; , \\
\d_{i}\t_{i} - q \t_{i}\d_{i} & = & 1 + (q-1)
\sum^{i-1}_{j=1}\t_{j}\d_{j} \;\; .
\end{array}
\ee
This algebra resembles the differential calculus on the quantum
hyperplane \cite{wz} (see also \cite{fz}). The correspondent graph
is a chain of $N$ arrows. Algebras of this kind can exist only for
even $p$, as mentioned above.

Concluding this discussion, we would like to formulate  some
problems related to complete classifying paragrassmann algebras.

1. It is clear that algebras $\Gamma_{p+1}(N)$ with different
sets $\{a_{i}\}$ (see (\ref{strip})) are not equivalent
(unless the two sets are proportional). The question is how
fully the {\em ansatz} (\ref{strip}) exhausts all admissible
matrices $\rho_{ij}$ in (\ref{m8})?
(We suspect that for $N$ large enough it is
exhaustive while for smaller $N$ it is not.)

2. An algebra $\P(N)$ can be determined by an oriented tree $\G$
together with a suitable set $\{a_{i}\}$ (or, more generally,
$\{\rho_{ij}\}$). Different trees and sets define
non-equivalent algebras that cannot be related by any linear
transformation of the variables $\t$ and $\d$. The question is: can they
be related by a {\em non}-linear transformation like that
connecting the versions of the algebra $\P(1)$?
In other words, can different $\P^{\{\G,a\}}(N)$ be considered
as versions of a unique algebra $\P(N)$?

The final remark concerns possible non-bilinear algebras.
Our approach can be  generalized
to arbitrary version with commutation relations (\ref{u7}).
With this aim, we first introduce a linear automorphism operator
$\gg$ in the algebra (\ref{u7}) that satisfies the commutation relations
(\ref{m6}) with some $q$, not necessarily a root of unity.
Then, for the multi-paragrassmann generators defined as in
Eqs.~(\ref{m7}), (\ref{m10}), one can derive the following algebra
($i<j$):
\be
\lb{i5}
\begin{array}{rcl}
\d_{i}\t_{i} & = &
b_{0} + b_{1} \t_{i}\d_{i} + b_{2}\t_{i}^{2}\d_{i}^{2} + \dots
+b_{p}\t_{i}^{p}\d_{i}^{p} \; , \\
\t_{i}\t_{j} & = & q^{a_{i}} \t_{j}\t_{i}    \; , \\
\d_{i}\d_{j} & = & q^{a_{i} } \d_{j}\d_{i}    \; ,     \\
\d_{i}\t_{j} & = & q^{-a_{i}}  \t_{j}\d_{i}  \; ,   \\
\d_{j}\t_{i} & = & q^{-a_{i}} \t_{i}\d_{j}    \; .
\end{array}
\ee
To satisfy the equations (\ref{m9}), (\ref{m13a}), we have chosen the
parameters $\sigma_{ij}=\rho_{ij} =a_{j}$ for $i<j$ and
$\sigma_{ij}=\rho_{ij} = -a_{i}$ for $i>j$. The integer numbers $a_{i}$
are restricted by the condition that all $q^{a_{i}}$ are primitive
$(p+1)$-roots of unity.
The most important feature of this construction is its independence
of the version (`version covariance'). This property is of utmost
importance in some applications, e.g. in constructing para-Virasoro
algebras to be treated in our next paper. Note, however, that the
generalized Leibniz rule (\ref{leibniz}) is only satisfied if the
$b$-parameters correspond to the $\gg-\bar{\gg}$--Version.
The algebra (\ref{i5}) may be further
generalized but we will not present these generalizations here.
Non-bilinear algebras deserve a separate thorough investigation.

\section{Conclusion}
\setcounter{equation}0
In this paper, we have given a general construction of the paragrassmann
calculus with one variable and have shown that all nondegenerate algebras
$\Pi_{p+1}^{\{b\}}$ are equivalent. Still, different versions may
be useful in different applications. As has been shown in the last
section, constructing algebras with many variables requires simplest
versions.  Another reason for a separate consideration of different
equivalent versions is the following. Our
approach to constructing paragrassmann calculus with many variables
was to preserve the nilpotency property for linear combinations
of $\theta_i$  and of $\d_i$. Then the commutation relations between
different elements are just calculational tools not having any
fundamental meaning. However, we may choose a different viewpoint,
considering the algebra of commutation relations as a prime object.
Then it would be natural to look for transformations preserving the
commutation relations.

Let us discuss this viewpoint. It is clear that transformations
\be
\lb{m24}
\d_{i} \rightarrow \d_{i}' = t_{ij} \d_{j} \; , \;\;
\t_{i} \rightarrow \t_{i}' = t_{ij} \t_{j} \; .
\ee
do not preserve the commutation relations (\ref{m8}) and (\ref{m11}).
To preserve these commutation relations (quantum hyperplane relations)
we have to consider $t_{ij}$ as generators of the multiparametric
quantum group $GL_{q,\rho_{ij}}$. In particular, we have to require
$$
t_{ik}t_{ij}=q^{\rho_{kj}}t_{ij}t_{ik} \; ,
$$
$$
\t_{k} t_{ij} = t_{ij}\t_{k} \; .
$$
The main paragrassmann identity now looks as
\be
\lb{m25}
(\t_{i}')^{p+1} = ( \sum^{N}_{j=1} t_{ij}\t_{j} )^{p+1} =0.
\ee
It is clear that
$$
(t_{ik}\t_{k})(t_{ij}\t_{j})
= q^{2\rho_{kj}} (t_{ij}\t_{j})(t_{ik}\t_{k}),
$$
and so Eq.~(\ref{m25}) is fulfilled only if $q^{2\rho_{ij}}$ are
primitive roots of unity. As an example we present the paragrassmann
quantum plane defined by
$ \t_{i}=\gg^{1/2} \otimes \cdots
\otimes \gg^{1/2}\otimes
\underbrace{\t \otimes 1 \otimes \cdots \otimes 1}_{i} $.
The nilpotency conditions $(\t_{i})^{p+1} =0$ are obviously satisfied
and the commutation relations,
\be
\lb{m26}
\t_{i}\t_{j} = q^{1/2} \t_{j}\t_{i}   \, , \;\; (i<j) \; ,
\ee
are not changed under the transformations (\ref{m24})
with $t_{ij} \in GL_{q^{1/2}}(N)$. Then it is clear that
$$
(t_{ik}\t_{k})(t_{ij}\t_{j})
= q (t_{ij}\t_{j})(t_{ik}\t_{k})
$$
and, if $q$ is primitive root of unity, we obtain that
$$
(\t_{i}')^{p+1} =0 .
$$
Thus, the paragrassmann quantum plane (\ref{m26}) may be regarded
as a linear space under rotations of the quantum group $GL_{q^{1/2}}$.

We hope that this remark shows a deeper connection between PGA and
quantum groups than suggested by simpler observations of Ref.~\cite{fik}.
It is quite possible that there exist other relations implicit in
 works on finite-dimensional representations of quantum groups with
a root-of-unity deformation parameter \cite{kac}.These representations
are very interesting  from the mathematical point of view \cite{kac}
and have recently found applications in rational conformal field
theories \cite{luis}. However, here and
now, we wish to emphasize that our motivation and our starting point
were quite far  from the theory of quantum groups and so the relations
to this field were somewhat surprising to us, especially, for general
algebras defined by Eq.~(7). In fact, our aim was to generalize the
Grassmann calculus and to apply the PGA to describing particles with
paragrassmann variables, fractional spin and statistics, para-conformal
and para-Virasoro algebras, etc. (the last topic is detailly treated in
our next paper \cite{fik1}). Up to now, the relations between PGA and
quantum groups, being themselves interesting and beautiful, were not
very helpful in these applications. We hope that better understanding
the nature of these relations might be useful for physics applications
both of quantum groups and of paragrassmann algebras. In this connection,
we have to stress that our fairly general construction of the
many-variable paragrassmann calculus is probably not the most general
one, and complete classifying of nondegenerate  algebras $ \P (N)$
is highly desirable.

One may hope that this eventually will open a way to applications of
PGA to systems of many particles with fractional spins and statistics.
It is conceivable that these applications  are not necessarily restricted
to physics in two spatial dimensions and that PGA might be applied to
some three-dimensional systems as well, e.g. to quarks inside hadrons.
The idea of considering confined quarks as soliton-like quasiparticles
has recently attracted some attention, and a model with one-dimensional
soliton-like quarks has been treated in detail \cite{ellis}. A further
step in this direction might be to look for anyon-like excitations
(vortices) on the surface of the hadron (the border between two phases of
the QCD). In this connection, recent results on braid-group analysis of
anyons on topologically nontrivial surfaces \cite{wu} might be of great
interest. Ref.~\cite{wu} clearly demonstrates that basic facts of the
anyon physics can be derived in terms of algebraic, topology-dependent
analysis and thus can be applied not only to strictly planar systems.
Of course, these remarks are highly speculative but
we decide to include them to hint at interesting physics
applications of our apparently abstract analysis.

\section{Appendix}
\setcounter{equation}0

Here we describe parafermions and parabosons [14] in the framework of
our approach to the paragrassmann algebras.

1. {\bf Parafermionic Version.} \\
Parafermionic generators $\t$ and $\d$ ($\t^{p+1}=\d^{p+1}=0$)
satisfy the commutation relations [14]
\be
\lb{a1}
[[\d, \; \t ] \t ] = - \rho \t \; , \;\;
[[\d, \; \t ] \d ] = \rho \d \; .
\ee
It is hard to extract the basis for the algebra with these generators,
because we can not move all $\d$'s to the right-hand side of any
monomial $\dots\d^{i}\t^{j}\d^{k}\t^{l}\dots$. Thus, our aim is to
find a structure relation (2.7) which is in agreement with
(\ref{a1}). To do this, we apply the relations (\ref{a1}) to the
vector $| k \rangle = (\t^{k})$. Then taking into account
Eqs.~(2.8) we derive the following recurrent equations
\be
\lb{a2}
\begin{array}{rcl}
\rho  =  2\a_{n} - \a_{n+1} -\a_{n-1} & , & n   = 1, \dots , p \; , \\
\a_{0} & = & \a_{p+1} \; = \; 0 \; .
\end{array}
\ee
The solution is
\be
\lb{a3}
\a_{n} = n\a_{1} - \frac{n(n-1)}{2}\rho \; , \;\;
\rho = \frac{2\a_{1}}{p} \; .
\ee
Choosing the normalization condition $\a_{1}=1$ we have
\be
\lb{a4}
\rho = 2/p \; , \;\; \a_{n} = n(p+1-n)/p \; .
\ee
 From Eqs.(2.10) one can find the parameters $b_{i}$
specifying the commutation relation for $\d$ and $\t$ (2.7).
For first few $b_{i}$ we obtain
\be
\lb{a5}
b_{0} =1\; , \; b_{1} = \frac{p-2}{p}\; , \; b_{2} = -\frac{2}{p(p-1)}\; ,
\; b_{3} = - \frac{4}{p(p-1)(p-2)}\; , \; \dots \; .
\ee
2. {\bf Parabosonic Version.}

This Version is specified by the commutation relations [14]
\be
\lb{a6}
[ \{\d, \; \t \} \t ] = \d\t^{2} - \t^{2}\d = - \rho \t \; , \;\;
[ \{\d, \; \t \} \d ] = \t\d^{2} - \d^{2}\t = \rho \d \; .
\ee
Now the recurrence equations are
\be
\lb{a7}
\begin{array}{rcl}
\a_{n+1} & = & \a_{n-1} - \rho \; , \;\; n=1, \dots , p \; ; \\
\a_{0} & = & \a_{p+1} = 0 \; .
\end{array}
\ee
A solution of these equations (for $\rho \neq 0$ ) exists
for even $p$ only. With $\a_{1}=1$ we obtain
\be
\lb{a8}
\begin{array}{rcl}
\rho & = & 2/p \; ,  \\
\a_{n} & = & -n/p \;\; {\rm for \; even} \; n \; , \\
\a_{n} & = & (p+1-n)/p \;\; {\rm for \; odd} \;  n \; .
\end{array}
\ee
As above, $b_{i}$ are derived by using Eqs.~(2.7)
\be
\lb{a9}
b_{0}=1\; ,\; b_{1} = -\frac{p+2}{2}\; , \; b_{2} = 2\frac{p+1}{p}\;,
\; b_{3} = 4\frac{p+1}{p(p-2)}, \; \dots \; .
\ee
Thus parafermionic and parabosonic algebras can also be defined
by the relations (2.3) and (2.7) with an appropriate
choice of the parameters $b_{i}$.

\section*{Acknowledgments}
A significant part of this work has been done while one of the
authors (A.T.F.) visited the Institute of Theoretical Physics of the
University of Turin, and the final version has been completed
in the Yukawa Institute of Theoretical Physics. A.T.Filippov wishes
to express his deep gratitude to V.~de Alfaro,  K.~Nishijima
and Y.~Nagaoka for arranging the visits and
for invaluable support in these hard times. Kind hospitality
and  the financial support of INFN and  of the YITP that
made completing this research possible are deeply appreciated.
Useful communications with A.D'Adda, M.Caselle, L.Castellani,
R.Floreanini, S.Forte, P.Furlan, A.LeClair, M.Mintchev, O.Ogievetsky,
V.Spiridonov, and I.Todorov are acknowledged by all the  authors.

\end{document}